# Weight Training Analysis of Sportsmen with Kinect Bioinformatics for Form Improvement

**Muhammad Umair Khan**[§]
umair.khan@kics.edu.pk

**Khawar Ali Shah**[§]
khawar.alishah@kics.edu.pk

**Sidra Qadeer**[§]

[§]Al-Khawarzimi Institute of Computer Sciences, UET, Lahore

## Abstract

Sports franchises invest a lot in training their athletes. Use of latest technology for this purpose is also very common. We propose a system of capturing motion of athletes during weight training and analyzing that data to find out any shortcomings and imperfections. Our system uses Kinect depth image to compute different parameters of the athlete's selected joints. These parameters are passed through certain algorithms to process them and formulate results on their basis. Some parameters like range of motion, speed and balance can be analyzed in real time. But for comparison to be performed between motions, data is first recorded and stored and then processed for accurate results. Our results depict that this system can be easily deployed and implemented to provide a very valuable insight to dynamics of a work out and help an athlete in improving his form.

**Keywords**

Motion Capture, Motion Analysis, Sports Training

## 1   Introduction

The sports industry is integrating more and more technology to improve the standards of the game. Use of high speed cameras, thermal cameras, vibration sensors, impact sensors and other such equipment is quite common in a number of sports for monitoring sportsmen themselves, and other parameters of the sport. Apart from the in-game analysis and observation, pre-game analysis is also of utmost importance and helps players to improve their techniques and have a better understanding of the controlling parameters of the sport.

Monitoring motion of player's body plays a fundamental role in correcting mistakes and improving performance in a lot of sports. Generally, the approach used for this is just rudimentary eye-balling: the coach has to keenly focus on player's motion and movements of his joints to see if anything looks off balance. Recently, research is being done to introduce certain technologies into this to get more accurate results. In this respect, a number of methods are used including IMUs (Inertial Measurement Unit), RGB and Depth Cameras. IMUs are sensors that can be attached to certain body parts and thus can calculate parameters like speed and position etc. Although, the data provided by IMUs is very accurate and has very low error values, but their implementation is quite arduous. They are often connected with wires to control unit that might get entangled and even restrict the motion of the user. Even if the IMUs are wireless, quite a lot of sensors are required to cover the whole body's motion and they require a powerful controller which can communicate with them simultaneously.

Motion capture with RGB cameras is also used in certain instances. This method uses an array of cameras at different heights that encircle the player. The image feeds of all these cameras is combined and processed to estimate the player's motion. This method turns out to be very costly in terms of hardware and processing.

Another type is motion capture through Depth Cameras. Depth cameras project a beam of IR dots and calculate distance of objects in environment based on time-of-flight of these dots. The dots bouncing off objects that are close to source have lesser time-of-flight than those that bounce off farther objects. This in return, builds a three dimensional image of the environment with millimeter precision. Now, image processing algorithms are applied to detect and isolate human form from the depth image and further processing can identify individual joints of the human being. Some arithmetic operations then calculate exact position in space, of each joint, its rotation and speed. This method is far more cost effective than motion capture with RGB cameras, both in terms of hardware and processing.

The proposed system is based on XBOX Kinect sensor which is equipped with a depth camera. Analysis is performed based on set standard values. The positions and angles of joints vary in a specific pattern for any kind of motion. We calculate the amount of error in relevant values for each frame. And devise a report based on error plots.

## 2  Previous Work

Researchers have major interest in understanding human kinematics. A huge amount of hardware and software tools have been developed so far for this purpose. Each, catering to their own capabilities and limitations. To be able to isolate minute details in motion of athletes in sports, importance of choice of hardware and software tool to be utilized is paramount. The type of sport in focus is also to be considered as dynamics of different sports may vary greatly.

Hassan Ghasemzadeh used a network of sensors attached to the body to measure rotational motion of wrist during golf swings [1]. Some researchers have integrated motion capture data into virtual reality to study the perception-action cycle of athletes. They tested their system on rugby players and found that VR provides a very interactive and manipulatable environment to help simulate a wide range of different scenarios. And thus, tends to provide much better insight relative to video playbacks [2]. Mathew Bodie et al. developed an IMU based system coupled with GPS to analyze biomechanics of skiers. They made a specialized suit with integrated wired connections and slots to allow the sensors to be directly attached to skin [3]. Fulan used multiple colour cameras, and with the help of certain tracking algorithms, created anatomically accurate models of human form. Comparison was performed, both quantitatively and qualitatively, with the help of a number of datasets. Mean absolute errors and standard deviations were formulated to see which methods fairs best [4]. Eline and Marco discussed accuracy of different motion capture systems for sporting applications. A number of parameters were tested including, team or individual play, indoor and outdoor environment and area of action etc. [5]. Another study focusing on combat sports and martial arts, presented several motion analysis systems that could be used for such sports. It highlighted the techniques best suitable for said application and resultantly, the benefit of using motion capture technologies in helping athletes and coaches [6]. Bernardina et al. used markers and an array of cameras to monitor underwater motion of an athlete's hand. Position of markers inside water was formulated according to reference coordinate system during different motions. They concluded that the proposed methodology could be used to monitor maneuvers of athletes in water-based sports like swimming, water polo and water aerobics etc. [7]. Experiments by Carlos Zerpa et al. provide evidence of reliability of the Microsoft Kinect system for measurement of human movement kinematics [8]. Alexiadis described a system that evaluates dance performances against a standard data set and provides visual feedback in virtual environment. Performer's motion data was acquired via Kinect based human skeleton tracking and performance scores were calculated for each dancer [9]. Hesham Alabbasi presented an approach to handle problems of medical rehabilitation and sport training by using Kinect sensor. Each exercise is first recorded as a model and then imitated by trainee. Comparison was performed based on difference between angles of different joints [10].

Our target being standard weight training and calisthenics, Kinect shows to be an appropriate choice. Kinect's raw depth data can be used to formulate a wide range of parameters like absolute positions, rotations and speeds. Also, Kinect's human and skeleton tracking can provide helpful visual aid and easily understandable representation of analysis results.

## 3  Skeleton Generation

The designed system uses Kinect for Windows SDK 2.0 and a set of specific libraries to extract raw data from the sensor. Upon the raw data, human detection algorithms are implemented and later on, location of specific joints in space is isolated. Some arithmetic operations are used to draw a skeleton of the subject in sensor's field of view. Kinect is capable of identifying a number of features in human body which are depicted in figure 1.

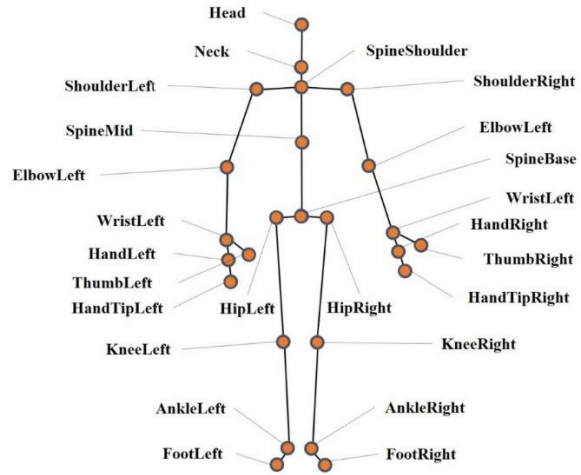

*Figure 1. Kinect Skeleton*

Each listed joint's position in terms of x, y and z coordinates can be extracted and manipulated in desired manner.

## 4  Features of Interest

Our system calculates a number of features from depth values of joints including range of motion, balance and alignment between paired joints, and position and speed magnitudes of selected joints. Range of motion of a limb or balance of an athlete's body during maneuvers is a key factor that defines an athlete's health and physical ability. Each feature has to be measured in separate manner.

### 4.1  Static Pose Matching

In this mode, as the user comes in sensor's field of view, the system starts calculating angles between joints of the devised skeleton. Position vectors are generated between the target joint and its adjacent upper and lower joints. Trigonometric operations are then applied to calculate angle between these vectors.

A single reference frame of choice is saved by the user and the system starts comparing angles of selected joints to those of the reference frame. Whenever user matches his/her pose exactly with the reference pose, the system generates a flag and notifies the user. This is particularly helpful if the athlete is trying to perfect a certain stance.

## 4.2 Range of motion

Range of motion is defined by the extent a certain part of body can be moved around. Ranges of motion of a healthy male adult, as defined by Nancy Hamilton is given in table 1 [11].

An athlete recovering from an injury can track his/her progress based on deviation from standard values. The user selects the joint to test and the system starts recording turning angles of that particular joint. Any variations from the standard are noted and an aggregate score is generated.

| Joint Name | Motion Type | Angle | Motion Type | Angle |
|---|---|---|---|---|
| Lumber Spine | Lateral Flexion | 35° | Hyper Extension | 20° |
| Elbow | Flexion | 140° | Hyper Extension | 10° |
| Shoulder | Abduction | 180° | Adduction | 50° |
| Shoulder | Flexion | 180° | Extension | 50° |
| Ankle | Dorsiflexion | 20° | Planar Flexion | 50° |

Table 1. Ranges of Motion for Average Adult

## 4.3 Balance

Maintaining balance during any motion in sports plays a key role in achieving optimum performance. During balance analysis mode, the system continuously compares the relative position of paired joints (e.g. Shoulders, elbows, knees etc.) in transverse and sagittal plane. Any imbalance and difference in height or depth of joints is noted and quantized.

## 4.4 Motion Comparison

This mode analyzes a string of motion by the user. For comparison, a reference stream of data has to be saved first. This mode proves to be helpful in helping an athlete or a coach in analyzing how the body moves during a particular action. It can provide minute details of orientation in space and other such features frame by frame.

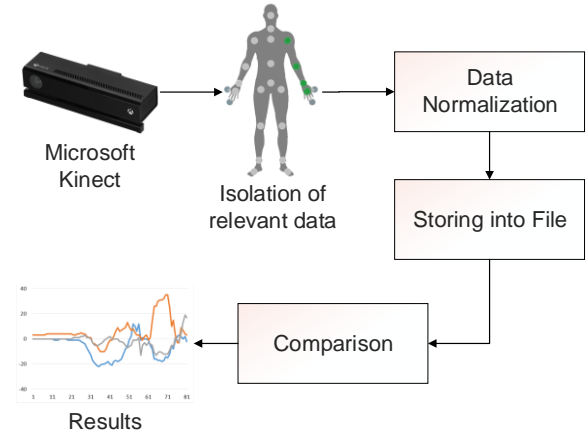

Figure 2. Process Flow

But for proper comparison of actions, data needs to be normalized. This is to avoid difference in values due to difference in height, body frame, relative position in space with respect to sensor and length of time taken to perform the action.

### 4.4.1 Standardizing User Height

To normalize height, a scaling factor is devised based on difference in heights of reference and test subjects. All position values of the test subject are normalized with respect to this element ($SF$).

$$SF = \frac{H_{ref}}{H_{test}} \qquad (1)$$

Here, $H_{ref}$ is height of the person performing reference pose and $H_{test}$ is height of the person performing test pose.

### 4.4.2 Standardizing Location in Field of View

To alleviate error due to difference in position of the subjects in our frame of reference, a new coordinate system is implemented with a particular point in body itself as the frame's origin.

### 4.4.3 Standardizing Number of Frames

The streams to be compared must be of the same length and synchronized. For this purpose, the test stream's number of frames is equalized with that of reference stream. In case of access of frames in test stream, appropriate number of frames are removed and in case of deficiency of frames, additional frames are added at particular places within the stream. The data for the added frames is interpolated from adjoining frames.

The normalized data values are then stored in a file for comparison. Error values are formulated between the reference values and test values. Results are deduced from plots of error values vs time for each relevant joint.

## 5 Position and Speed Processing

The raw data provided by Kinect doesn't itself provide very clean values. Some filters and smoothing functions have to be applied for adequately analyzable figures.

These issues and noise arises due to intrinsic limitations of skeleton tracking through Kinect. Even if a subject is in stand still position, the skeleton appears to be somewhat flickering or vibrating at certain points. This introduces a lot of ripples in the readings. For the data to be meaningful to an average user, it has to be smoothed out. Applying a second order moving average filter on raw data brings very reasonable results. A moving average filter calculates average of elements within a fixed window while the window keeps shifting forward and can be represented in the form of equation below.

$$x_t = \frac{1}{2n+1} \sum_{i=-n}^{i=n} x_{t+i} \qquad (2)$$

Experiments were performed with different athletes and novice subjects. Actions performed by athletes were saved as reference data and novice subjects were asked to perform the same actions. There actions were recorded and compared.

Figure 4 shows error in position of a subject's hand while performing a bicep curl. It can be concluded from the chart that the subject is pulling his hand too high up during first half of the maneuver and bringing it too low during the second half. While the error in the hand's lateral movement is within tolerance limit.

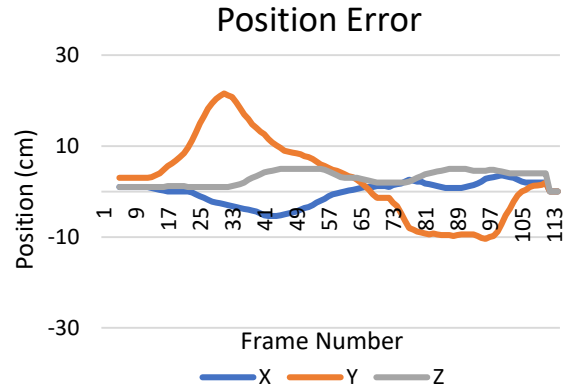

*Figure 3. Position Error in Subject's Hand during Bicep Curls*

Another subject was instructed to perform a push press. Figure 5 shows a comparative graph of speed of that subject's elbow. It can be observed from the chart that the subject's average speed is quite similar to reference speed while pushing the barbell up. But his acceleration is off point while bringing it back down towards his shoulder.

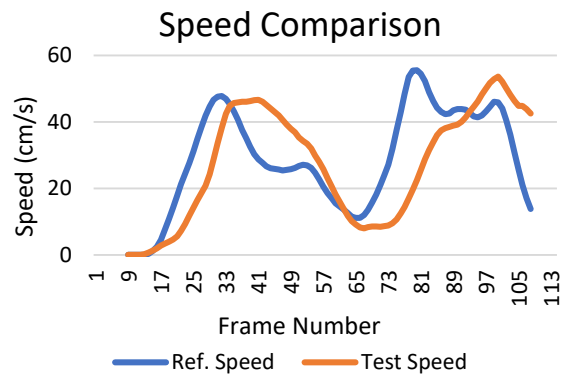

*Figure 4. Comparison of Speed of hand during a Push Press*

Such in-depth information is quite vital during training and provides a quantitative measure of deficiency or improvement in users.

# 6  Performance Score

Performance score (PS) is devised as a function of weighted averages of position errors, speed errors and degree of imbalance.

A cumulative error of x, y and z coordinates and speed of all relevant joints is calculated for each frame which is then averaged over the whole length of time.

$$PS = \frac{w_p E_p + w_s E_s + w_b E_b}{3} \qquad (3)$$

Here, $E_p$, $E_s$ and $E_b$ represent average position error, speed error and balance error, and $w_p$, $w_s$ and $w_b$ represent their corresponding weights respectively.

# 7  Results

The positions of various joints, angles between them, and the speed of motion are among the deciding factors of efficiency of a work out. A person who isn't taking care of all these factors most often than not, ends up only wasting a lot of energy and time with minimum gains. And in some worse cases, even injuring himself/herself.

To test whether the devised system helps improve work out efficiency, a number of experiments were conducted with various subject of varying experience in weight training. For reference data, a pro athlete was asked to perform some specific exercises and the data from all these runs was recorded and saved. All other subjects were asked to perform the same exercises and their data was then analyzed. The performance measure score for each individual turned out to be in line with the subject's level of experience. Those, more experienced in those exercises had smaller cumulative errors and novices returned larger cumulative errors. Each individual had to perform the exercises multiple times, each time after looking at their respective results put forth by the system. The numbers indicate a learning curve from most subjects as they improve their form in each run.

| Subject No. | Performance Score | | | | |
|---|---|---|---|---|---|
| | Run 1 | Run 2 | Run 3 | Run 4 | Run 5 |
| 1 | 15.3 | 15.8 | 13.1 | 13.5 | 12.6 |
| 2 | 25.2 | 26.7 | 22.9 | 18.5 | 17.8 |
| 3 | 40.7 | 32.2 | 30.1 | 15.9 | 10.4 |
| 4 | 10.6 | 11.3 | 9.8 | 9.0 | 9.1 |
| 5 | 17.3 | 12.9 | 8.5 | 6.2 | 6.3 |

*Table 2. Performance Scores of Test Subjects (Lower value is better)*

Table 2 shows performance measure of five subjects performing a bench press. The error clearly decreases gradually by fifth run as the subjects learn from their mistakes and apply corrections. Since the performance measure is a function of cumulative errors, it correctly reflects experience and expertise of each subject.

# 8  Conclusion

We have presented a non-invasive, marker-less, vision-based method to analyze different actions in weight training to help athletes improve and perform their best. Our proposed system can monitor a number of parameters, discussed previously, and output them as per requirement. We have shown how to manipulate the raw data from Kinect and make sense of it relative to particular actions preformed.

For future work, we can work on relinquishing the need to store data for analysis to save processing cost and be able to generate results in real time. To reduce ripples and noise in raw data, multiple Kinect sensors can be added at different locations and their correlation drawn, thus generating a much smoother raw data.

Kinect motion capture proves to provide valuable insight on sports bioinformatics and has a potential of becoming a rudimentary asset in future sports industry.